\documentclass{svjour3}
\usepackage{hyperref,graphicx,stmaryrd,amsmath,amssymb}
\usepackage[latin1]{inputenc}
\newcommand{\mb}[1]{\mathbf{#1}}
\newcommand{\mr}[1]{\mathrm{#1}}
\newcommand{\mmu}{{\mbox{\boldmath$\mu$}}}
\newcommand{\nnu}{{\mbox{\boldmath$\nu$}}}
\newcommand{\ssigma}{{\mbox{\boldmath$\sigma$}}}

\newcommand{\bna}{\boldsymbol{\nabla}}

\newcommand{\Fpl}{\mb{F}_{\!\mr{p}}}
\newcommand{\Fel}{\mb{F}_{\!\mr{r}}}
\newcommand{\Fmmu}{\mb{F}_{\!\mu}}
\newcommand{\FEPN}{\mb{F}_{\!\text{p}}^G}
\newcommand{\Spm}{\mb{S}}

\newcommand{\psitot}{\psi}
\newcommand{\psiel}{\psi_{\mr{r}}}
\newcommand{\psipot}{\psi_{\mu}}

\newcommand{\GLZ}{\mr{GL}(3,\mathbb{Z})}
\newcommand{\curl}{\mathop{\mathrm{curl}}}
\usepackage{xcolor,ulem,cancel}

\journalname{Journal of Elasticity}

\begin{document}

\title{Intermittency in crystal plasticity informed by lattice symmetry}

\author{Paolo Biscari \and Marco Fabrizio Urbano \and Anna Zanzottera \and Giovanni Zanzotto}

\institute{P.~Biscari and A.~Zanzottera \at Department of Physics, Politecnico di Milano, Piazza Leonardo da Vinci 32, 20133 Milano, Italy \email{paolo.biscari@polimi.it} \and M.F.~Urbano \at SAES Getters, Viale Italia 77, 20020 Lainate, Italy \and G.~Zanzotto \at DPG, Universit\`a di Padova, Via Venezia 8, 35131 Padova, Italy}

\maketitle

\begin{abstract}
We develop a nonlinear, three-dimensional phase field model for crystal plasticity which accounts for the infinite and discrete symmetry group $G$ of the underlying periodic lattice. This generates a complex energy landscape with countably-many $G$-related wells in strain space, whereon the material evolves by energy minimization under the loading through spontaneous slip processes inducing the creation and motion of dislocations without the need of auxiliary hypotheses. Multiple slips may be activated simultaneously, in domains separated by a priori unknown free boundaries. The wells visited by the strain at each position and time, are tracked by the evolution of a $G$-valued discrete plastic map, whose non-compatible discontinuities identify lattice dislocations. The main effects in the plasticity of crystalline materials at microscopic scales emerge in this framework, including the long-range elastic fields of possibly interacting dislocations, lattice friction, hardening, band-like vs.~complex spatial distributions of dislocations. The main results concern the scale-free intermittency of the flow, with power-law exponents for the slip avalanche statistics which are significantly affected by the symmetry and the compatibility properties of the activated fundamental shears.
\end{abstract}

\section{Introduction}

The high-resolution investigation of plastic flow in crystalline materials at microscopic scales shows, in contrast to its smooth macroscopic behavior, a bursty dynamics related to the nucleation and motion of lattice dislocations, with power-law statistics for the slip avalanches \cite{dim06,08bri,uch09,zha12,gu13}. Theoretical work on these phenomena has been developed by using discrete dislocation dynamics \cite{02mig,08dev}, pinning-depinning \cite{04mor}, or automaton models \cite{sal11,sal12}, renormalization group methods \cite{09dah}, and continuum descriptions based on Nye's dislocation density tensor \cite{09fre}. Phase field methods have also provided fruitful tools for the analysis of many aspects of crystal plasticity \cite{wan01,wan03,rod03,kos04,kosor04,she04,wan10,levi12,levi13AM,levi13IJP,mam14,levi14}.

As defect nucleation, annihilation, and movement originate from concentrated slip processes which accommodate large strains while locally striving to preserve the original lattice structure, here we propose a new nonlinear three-dimensional (3D) phase field model accounting for crystal symmetry.
This is described by the global symmetry group $G$ of crystals, conjugated to the group $\GLZ$ of unimodular $3\times3$ integral matrices preserving 3D Bravais lattices \cite{eri80,bha04,con04,ari05,PZbook}. The group $G$ dictates the location and distribution of countably-many relaxed states in the space of strain tensors (see also a related approach in \cite{denoual10}). In this framework the plastic distortion map $\FEPN$ is therefore discrete as it is $G$-valued, parameterizing the energy well locally visited by the material's deformation gradient $\mb{F}$ at any $\bf{x}$ and $t$. The discrete jumps in the plastic map $\FEPN$ are related to the presence of lattice dislocations in the material, whose long range elastic fields agree with the classical linear elastic predictions, emerge as non-compatible discontinuities of $\FEPN$.

Near any $\FEPN \in G$ the amount of shear along each principal slip direction of the crystal constitutes a scalar phase field component. These fields select the deformation path through energy minimization, and the coupling with the discrete plastic distortion allows the material system to explore the complex energy landscape induced by crystal symmetry. The evolving phase fields and body deformation lead in turn to a crystallographically-informed dislocation dynamics, driven by energy relaxation, which automatically accounts for the possible presence of simultaneously activated, interacting multiple slip modes, without the addition of supplementary hypotheses.

The model can resolve isolated defects in the crystal as well as configurations possibly involving complex defect topologies. By simulating the behavior of the material under cyclic mechanical loading, we show that this framework describes correctly various features of the plastic flow in crystalline materials, and sheds new light on the properties of the associated scale-free intermittent behavior at microscopic scales.

The paper is organized as follows. In Sect.~\ref{sec:freeen} we describe the phase field model, including the free-energy functional driving the gradient-flow dynamics, together with the new discrete flow rule governing the evolution of the plastic distortion $\FEPN$. In Sect.~\ref{sec:flow} we evidence how $\FEPN$ characterizes dislocations within the material, and we test the model by analyzing numerically the mechanical response of a single crystal subject to different types of cyclic loadings. We also study some of the statistical properties of the bursty dynamics emerging in these simulations. We draw the main conclusions in Sect.~\ref{sec:disc}.

\section{Model}\label{sec:freeen}

Let an initially homogeneous stress-free single crystal occupy a reference domain $\Omega$ in 3D space. We denote by $\mb{y}$ the displacement field, with $\mb{F}=\bna \mb{y}$ the deformation gradient, $J=\det\mb{F}$, and, respectively $\mb{C}=\mb{F}^\top{\mb{F}}$ and $\mb{E} = \frac12(\mb{C} - \mb{I})$, the Cauchy-Green and strain tensors. Let $\mb{e}_j =  {e^h}\!_j \mb{u}_h$ (summation understood, with $j,h = 1, 2, 3$) be a basis for the reference crystal lattice, where $ \mb{u}_j$ are pre-set mutually-orthogonal unit vectors. The global symmetry group is then given by $G =  E^{-1}\GLZ  E$, with $E=({e^h}\!_j)$, see \cite{PZbook}.

Further information regarding the crystallographic features of plastic flow in the material is given by the $p$ principal slips, defined by the shears  $\Spm_\alpha = \mb{I}\pm\mb{b}_\alpha\otimes \mb{n}_\alpha \in G$, mapping the crystal to itself ($\alpha=1,\dots,p$). The Schmid tensors $\mb{b}_\alpha\otimes \mb{n}_\alpha$ are  obtained from mutually orthogonal vectors $\mb{b}_\alpha,\mb{n}_\alpha$, giving respectively the $\alpha$-th slip direction and slip plane normal. These are dictated by the crystallographic and mechanical properties of the material.

We assume the standard Kr\"oner-Lee decomposition of the deformation gradient $\mb{F}=\mb{F}_\text{\!e}\mb{F}_\text{\!p}$ in terms of an elastic and a plastic component, where $\mb{F}_\text{\!p}$ parameterizes the relaxed states of the material \cite{gurtin,conti14}. As the latter are given here by the tensors in $G$, we are led to the natural assumption that $\Fpl$ coincides with the discrete field $\FEPN \in G$, defined as the bottom of the $G$-related energy well to which the gradient $\mb{F}$ belongs at any given place $\bf{x}$ and time $t$. We further assume the elastic distortion $\mb{F}_\text{\!e}$ to be given by a component $\Fmmu$, accounting for the activated principal lattice slips, multiplied by a possible further residual distortion $\Fel$. Explicitly, we set
\begin{equation}
\label{FPL0}
\mb{F}=\mb{F}_\text{\!e}\mb{F}_\text{\!p}=\Fel\Fmmu\FEPN,
\end{equation}
with
\begin{equation}
\label{FPL1}
\mb{F}_\text{\!p}=\FEPN \in G \qquad \text{and}\qquad
\Fmmu=\mb{I}+\sum_{\alpha=1}^p \phi(\mu_\alpha)\,\mb{b}_\alpha\otimes \mb{n}_\alpha,
\end{equation}
where the variables $\mu_\alpha \in [-1,1]$ measure the amount of shear along each $\Spm_\alpha$.
As in \cite{levi12}, the role of the function $\phi(s)=3s|s|-2s^3$ in (\ref{FPL1})$_2$, which is odd and monotonically increasing in [$-1,1$], with stationary points in $0$ and $\pm1$, is to locally extend the stability range of all the $G$-related strain equilibria considered below.

Due to assumptions (\ref{FPL0})-(\ref{FPL1}), here the variables $\mu_\alpha$ act as phase fields interacting with the deformation field $\mb{y}$. This is further implemented by the assumptions we make for the anisotropic free energy density $\psitot$ of the material, which is taken to have the form $\psitot=\psipot + \psiel$, where the phase field related term $\psipot$ has the piecewise-polynomial expression
\begin{equation}
\label{eq:psipot}
\psipot(\mmu)=A_1\sum_\alpha\mu_\alpha^2\big(1-|\mu_\alpha|\big)^2+A_2\sum_{\alpha\neq \beta}\mu_\alpha^2\mu_\beta^2,
\end{equation}
with $\mmu=(\mu_1,\dots,\mu_p)$, while $\psiel$ gives an isotropic penalty, with elastic modulus $K$, to the residual distortion $\Fel$:
\begin{equation}\label{eq:strainen}
\psiel=\textstyle\frac{1}{2}\,K\left|\Fel^\top\Fel-\mb{I}\right|^2.
\end{equation}
For $A_h>0$, $\psipot$ is minimized when at most one among the $\mu_\alpha$ is equal to the transition value $\pm1$, and the other vanish \cite{stein96}. The potential $\psipot$ enforces the required symmetry under permutations of the $\mu_\alpha$, and also implies the symmetry with respect to the backward and forward motion through the principal shears. The $A_2$-term favors the activation of one single slip at a time. The ratio $A_2/A_1$ determines the width of the optimal shearing paths close to the coordinate axes whereon the system moves from one minimizer to another, as the greater $A_2$ the smaller the remaining phase fields may be when one of them departs significantly from zero. Finally, the ratio $A_1/K$ controls the height of the energy barriers between slip-related minimizers (and therefore the plastic yield stress). The adopted minimal expressions for $\psiel$ and $\psipot$ model two dominant effects occurring in plastic shear, i.e.~the presence of (tunable) energy barriers, and the preference for single slips. A more precise characterization of $\psitot$ would allow to mimic the detailed structure of the energy landscape of crystals within each well, as with anisotropic $\gamma$-surfaces given for instance by \textit{ab-initio} models \cite{mil08,cha14,dez14}. However, we keep the constitutive choices in the present framework as simple as possible\footnote{\label{foot1} As for instance in \cite{denoual10}, we do not include here any interface-penalizing gradient term in the free-energy functional because in all our simulations we keep the typical distance $H$ between interpolation nodes always greater than or equal to the minimal physically relevant length $h$ in the continuum, which in the present context is taken of the order the lattice cell spacing. To model plastic behavior of crystals at scales $H$ larger than $h$, besides including a gradient term in the phase field potential, the Schmidt tensors in \eqref{FPL0} must also be suitably re-scaled (see for instance \cite{levi12}). The corresponding scaling imposed on the potential $\psipot$ determines whether in the limit $H\gg h$ the yield and Peierls stresses vanish or stay finite. In what follows we only consider plasticity at microscales with $H\simeq h$.}

The evolution of the pair $(\mb{y},\mmu)$ is determined by a gradient-flow equation of the Ginzburg-Landau type, coupled with the standard Cauchy equation for mechanical equilibrium
\begin{equation}
\label{S1}
\frac{\partial \mmu }{\partial t}=-L\,\frac{\delta \psitot}{\delta \mmu}, \qquad
\mbox{div}\, \ssigma=\mb{0},
\end{equation}
where $\ssigma=J^{-1}\mb{F}\big(\partial_\mb{C} \psitot\big)\mb{F}^\top$ is the Cauchy stress. As with (\ref{eq:psipot})-(\ref{eq:strainen}), to limit the number of constitutive parameters in the model we consider in (\ref{S1}) an isotropic mobility coefficient $L$ multiplying the variational derivative of $\psitot$ with respect to $\mmu$. Equations \eqref{S1} are solved subject to initial conditions for $\mmu$, and to either displacement or traction boundary conditions for $\mb{y}$ on $\partial\Omega$.

Along the evolution of $\mmu$, the map $\FEPN$ must be discretely updated at any $t>t_0$ when one of the phase fields, say $\mu_\alpha$, reaches the value $\pm1$, as this indicates the local attainment by $\mb{F}$ of a $G$-neighboring relaxed state under the driving:
\begin{equation}
\label{updateFG}
\FEPN\to (\mb{I}\pm\mb{b}_\alpha\otimes \mb{n}_\alpha)\FEPN,
\end{equation}
the corresponding $\mu_\alpha$ being reset to zero. Rule (\ref{updateFG}) provides a tensorial generalization of the scalar updating considered in previous simplified models of crystal mechanics \cite{PRL_GZ,sal11,sal12}. The remaining phase fields, other than $\mu_\alpha$, are updated in such a way to guarantee the continuity of the combination $\Fmmu\FEPN$ in \eqref{FPL0} across jump instants. With $\mb{F}$ continuous by hypothesis, the above means that $\Fel$, $\psitot$, $\psipot$, $\psiel$ and $\ssigma$ are also always continuous, despite the discontinuities inherent in the evolution of the discrete map $\FEPN$ in (\ref{updateFG}) and in the corresponding updated $\mu_\alpha$. An exception to this general behavior occurs when, during the evolution of $\mmu$, the resulting $\Fmmu$ in (\ref{FPL1})$_2$ is not isochoric. For instance, this happens when
for some $\alpha,\beta$, two shears $\Spm_\alpha$ and $\Spm_\beta$ are active in (\ref{FPL1})$_2$ at some jump instant, with $(\mb{b}_\alpha\cdot\mb{n}_\beta)(\mb{b}_\beta\cdot\mb{n}_\alpha)\neq 0$ and non-zero $\phi(\mu_\alpha)$ and $\phi(\mu_\beta)$. In this case, if $\mu_\alpha$ reaches a transition value $\pm1$, then $\mu_\beta$ is set equal to zero to recover the natural isochoricity assumption on $\Fmmu$. The resulting discontinuity of $\Fel$ may be kept under control, as it becomes smaller the larger we choose $A_2$ in $\psipot$. The cross term proportional to $A_2$, indeed, penalizes non-zero values for any $\mmu$ components when any of them approaches $\pm1$.

The discrete updating rule (\ref{updateFG}) for $\FEPN$ has here the role that is played in the classical continuum theories of plasticity \cite{asaro,04boce,gurtin} by the standard differential flow rule for $\Fpl$, which is used also in phase field models for martensitic transformations and plasticity in crystals\cite{levi12,levi13AM,levi13IJP,levi14}. However, (\ref{updateFG}) represents a significant novelty of the present approach, as it enforces the condition that the material locally reaches a stress-free equilibrium through (\ref{S1})-(\ref{updateFG}) only if $\mb{F}$ belongs to $G$ modulo an orthogonal transformation.
This is because, in an unperturbed system $\mb{F}$ is a solution of
(\ref{S1}) only when both the phase field vector $\mmu$ minimizes $\psipot$ (and therefore, from (\ref{FPL0})-(\ref{FPL1}), $\Fmmu\FEPN\in G$), and $\Fel$ is orthogonal, as required. In previous phase field implementations of crystal plasticity (either in the linear \cite{wan01,wan03,rod03,kosor04,she04,mam14} or in the finite-strain \cite{levi12,levi13AM,levi13IJP,levi14} regimes), the relaxed states obtained by using the classical differential flow rule result to be lattice invariant maps only when a single slip is active because only in this case integer values of the phase fields correspond to tensors in $G$. By contrast, in the general case the values of $\Fpl$ obtained from the classical flow rule bear no relation to the underlying lattice structure of the crystal expressed by $G$.

\section{Results}\label{sec:flow}

\textbf{\bf Dislocations.~}The discrete field $\FEPN$ has a crucial role in locating dislocations, which are associated with its non-compatible spatial jumps. We recall that in classical continuum theories \cite{01cergur} the plastic distortion map $\Fpl$ describes a dislocation whenever it lacks compatibility, i.e.~when it is not possible to write it as the gradient of a vector field. In the present setting based on the discrete plastic distortion $\FEPN$, dislocations may occur in regions where the map $\FEPN$ is not compatible (analytically, where $\curl\FEPN \neq\mb{0}$, though this should be intended in the  distributional sense as $\FEPN$ is a piece-wise constant field). As a typical instance, we consider a partition of the reference domain $\Omega$ into subsets on which $\FEPN$ is constant. The induced jump discontinuities of $\FEPN$ are compatible if they satisfy the Hadamard conditions $\llbracket\FEPN\rrbracket=\mb{b}\otimes\nnu$, with $\nnu$ the unit normal to the jump surface \cite{PZbook}. This in particular implies that any compatible discontinuities must necessarily occur along (portions of) planar surfaces. When two such parallel surfaces, together with a further surface connecting them (which is necessarily non-compatible), are considered on a small neighborhood of a line, this latter is identified as a dislocation line in the lattice, so that compatibility failures -- including the presence of non-straight portions of jump surfaces -- may induce surface dislocations \cite{bil64,cer94,ach07}.

An explicit example illustrates how dislocations are described by the non-compatible discontinuities of $\FEPN$. We consider a simple-cubic crystal in $\Omega$, with mutually-orthogonal lattice vectors $\mb{e}_j = a \mb{u}_j$, $j = 1, 2, 3$, where $a$ is the lattice parameter, and $\mb{u}_j$ are unit vectors along the 4-fold axes. In this lattice there are $p=6$ fundamental slips $\Spm_\alpha$, which involve all the mutually-orthogonal couples of vectors $\mb{u}_j$. In this case the $\Spm_\alpha$ generate multiplicatively the entire symmetry group $G$, which for this Bravais type coincides with $\GLZ$.

\begin{figure}
\centering
\includegraphics[width=.95\textwidth, clip, keepaspectratio]{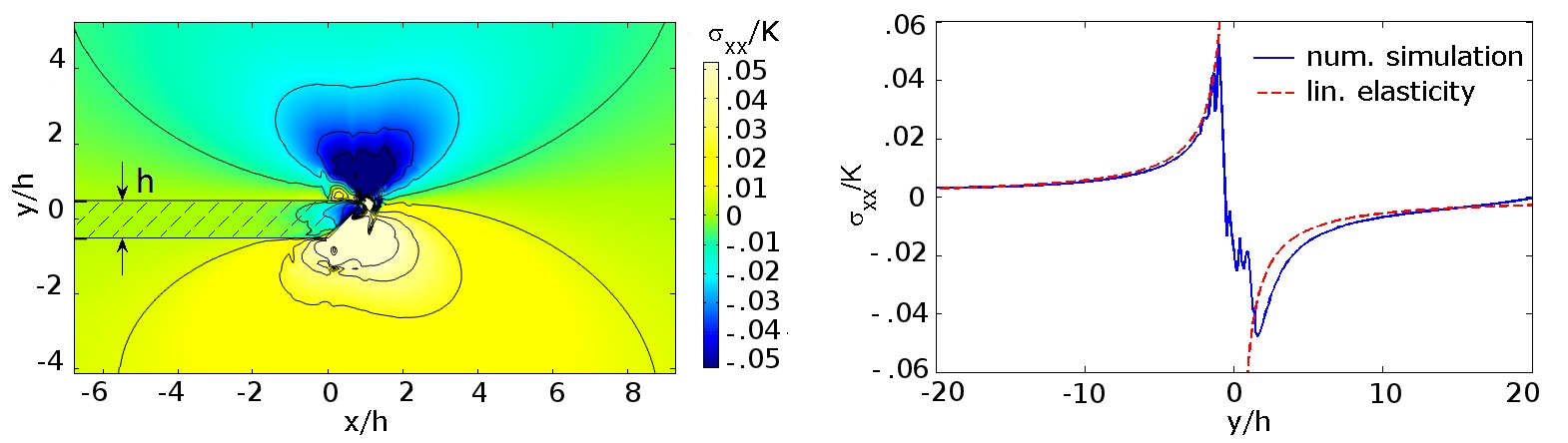}
\caption{(Color online) Left: Computed stress field near a dislocation. Right: Comparison of the results obtained for $\sigma_{xx}$ as a function of the distance from the center of the dislocation, for $x=0$: present model (solid line) and linear-elastic solution (dashed line).}
\label{fig:stressdistance}
\end{figure}

As in Fig.~\ref{fig:stressdistance}, we select $\Omega'\subset \Omega$ with boundary coinciding with the jump surface for $\FEPN$, composed by two nearby planes orthogonal to $\mb{e}_2$, with distance $h$ (see footnote~\ref{foot1}) and connected by a (necessarily non-compatible) portion of surface close to a (dislocation) line $\gamma_\text{dis}$. By considering now a closed curve $\gamma$ in $\Omega$ intersecting the jump surface in two points $P_1$ and $P_2$, one on each parallel plane, with $P_2=P_1+h\,\mb{e}_2$, we compute $\oint_\gamma \FEPN\mb{t} \,dS=\llbracket\FEPN\rrbracket(P_1P_2)=h\,\mb{b}$,
with $\mb{t}$ the unit tangent along $\gamma$. Therefore, in this case $h\mb{b}$ is the classical Burgers vector associated with the dislocation \cite{gurtin}. The standard edge and screw dislocations are retrieved as non-compatible discontinuities of $\FEPN$ for which $\mb{b}$ is respectively orthogonal or parallel to the line $\gamma_\text{dis}$.

To compare the stress field computed for a dislocation with the classical results derived from linear elasticity, we set $\FEPN$ to be respectively equal to $\mb{I}$ and $\mb{I}+\mb{u}_1 \otimes \mb{u}_2$, in $\Omega\backslash \Omega'$ and $\Omega'$. An edge dislocation results in $\gamma_\text{dis}$, with Burgers vector $h\mb{u}_1$. The $\sigma_{xx}$-component of the corresponding stress field is plotted in the left panel of Fig.~\ref{fig:stressdistance}, computed with the Comsol - Finite Elements Method software, used hereafter in all numerical simulations. In the right, this solution is compared with the analytical prediction \cite{hirth}, with shear modulus and Poisson ratio dictated by the potential \eqref{eq:strainen}. We see that the numerical simulation matches the analytical prediction to distances of the order of the defect width $h$.

Starting from a local minimizer identifying a lattice dislocation as in Fig.~\ref{fig:stressdistance}, the Peierls stress $\sigma_\text{P}$ is given by the external load at which the defect first becomes unstable, and here can be computed by considering a monoparametric family of equilibria of the system where all the phase fields are quenched except for the one associated with the principal slip active in $\Omega'$.
The value of $\sigma_\text{P}$ depends on all the energy constants, and by setting (here and in all the ensuing numerical simulations) the elastic modulus to $K=200\,$GPa, with $A_1/K=1$, $A_2/K=5$, and $h = 1\,$nm, we obtain $\sigma_\text{P} \sim 1.6\,$GPa, in line with values computed in other theories of crystal plasticity \cite{07rod,09rod}.

{\bf Plastic flow under cyclic loading.~}We now study the response of a single crystal to a possibly cyclic loading to show explicitly how the creation and evolution of dislocations spontaneously progresses during plastic flow in this model. We consider a system initially in local equilibrium, with vanishing phase field components, and $\Fpl = \FEPN$. The well-tracking map $\FEPN$ is initially set to coincide with the identity, except for the presence of possible quenched defects in the body, modeled by randomly distributed small regions where the phase field components and the map $\FEPN \neq\mb{I}$ are not allowed to evolve. We impose five different loadings I-V, on a simple-cubic crystal as above, occupying a box domain $\Omega$ with free lateral sides. The bottom of $\Omega$ has unit normal $\mb{m}$, and the lattice basis $\mb{e}_j$ is along the 4-fold axes. In simulation I the basis $\mb{e}_j$ is aligned with all the sides of $\Omega$, while in simulation II the crystal is rotated around the bottom normal $\mb{m}$ by about 45 degrees. The boundary conditions fix the bottom of $\Omega$ and impose a top-face deformation consistent with an overall shear $\mb{I}+ \lambda\,\mb{a}\otimes \mb{m}$,  where the vector $\mb{a}$ is orthogonal to $\mb{m}$ and parallel to one of the free box faces. These boundary conditions are designed to activate mostly one or two (compatible) slip systems during plastification. In simulation III the same boundary conditions are imposed but the lattice basis in $\Omega$ is rotated by about 30 degrees around the vector $\mb{a}$, so as to possibly activate more than two compatible slip system in the lattice. The more general loadings IV and V are then imposed to activate mostly non-compatible slips. Specifically, in case IV the isochoric deformation on the top and bottom faces of $\Omega$ is consistent with $\mb{I}+ \lambda(\mb{e}_1\otimes \mb{e}_3 + \mb{e}_3\otimes \mb{e}_2)$, while in simulation V an arbitrary rotation is considered of crystal axes $\mb{e}_j$ in the box, with the same imposed boundary conditions as in cases I, II, III. The parameter $\lambda$ is driven quasi-statically, so that the system evolves by local minimization through inhomogeneous equilibrium states, obtained by solving the equations in \eqref{S1}  coupled with \eqref{updateFG}. The resulting deformation fields are expected to activate a single principal slip in simulation I, and (mostly) two kinematically-compatible principal slips in simulation II (thus allowing in principle for defect-free interfaces). In contrast, in simulations III, IV and V, several slips are activated, possibly compatible or non-compatible among each other. In all cases the map $\FEPN$ tracks where and how such slips progress and eventually interact in the material, with the finite and multiplicative updating rule \eqref{updateFG} for $\FEPN$ guaranteeing the relaxed configurations to locally be driven to those dictated by lattice invariance.

\begin{figure}[t!]
\centering
\includegraphics[width=.95\textwidth, clip, keepaspectratio]{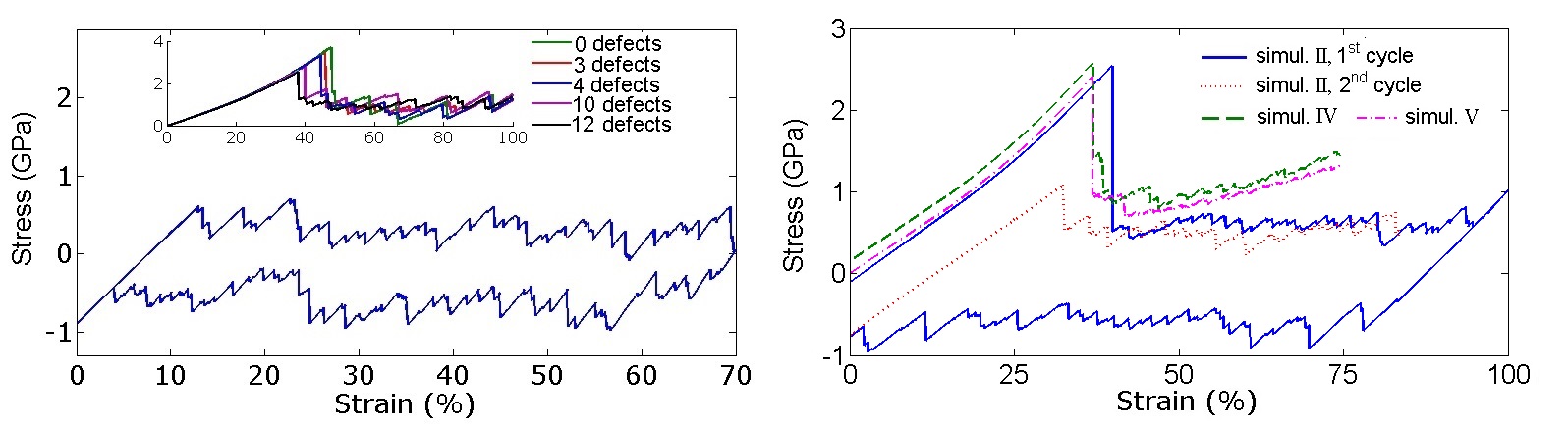}
\caption{(Color online) Intermittent plastic flow under cyclic shearing, in a box domain $\Omega$ with two edges parallel to $\mb{a}$ and $\mb{m}$, containing few randomly-positioned quenched defects. Relevant components of the stress and strain tensors are monitored. Left: Serrated stress-strain hysteretic loop for a typical cycle of loading-unloading with a single active slip (simulation I), and no hardening. Inset: decrease of the stress peak on the first loading, for an increasing number of identical defects randomly distributed in the sample. Right: Stress-strain curves in other simulations. Loading II (continuous blue and red dotted lines) mostly involves two compatible activated slips. No hardening is observed as for case II. The results of simulation III (not shown) also involve mostly compatible slips and produce no hardening, analogously to cases I and II. Simulation IV (green), and simulation V (purple) mostly involve non-compatible slips and show a clear hardening effect.}
\label{fig:FEPN_hyst}
\end{figure}

\begin{figure}[t]
\centering
\includegraphics[width=.95\textwidth, clip, keepaspectratio]{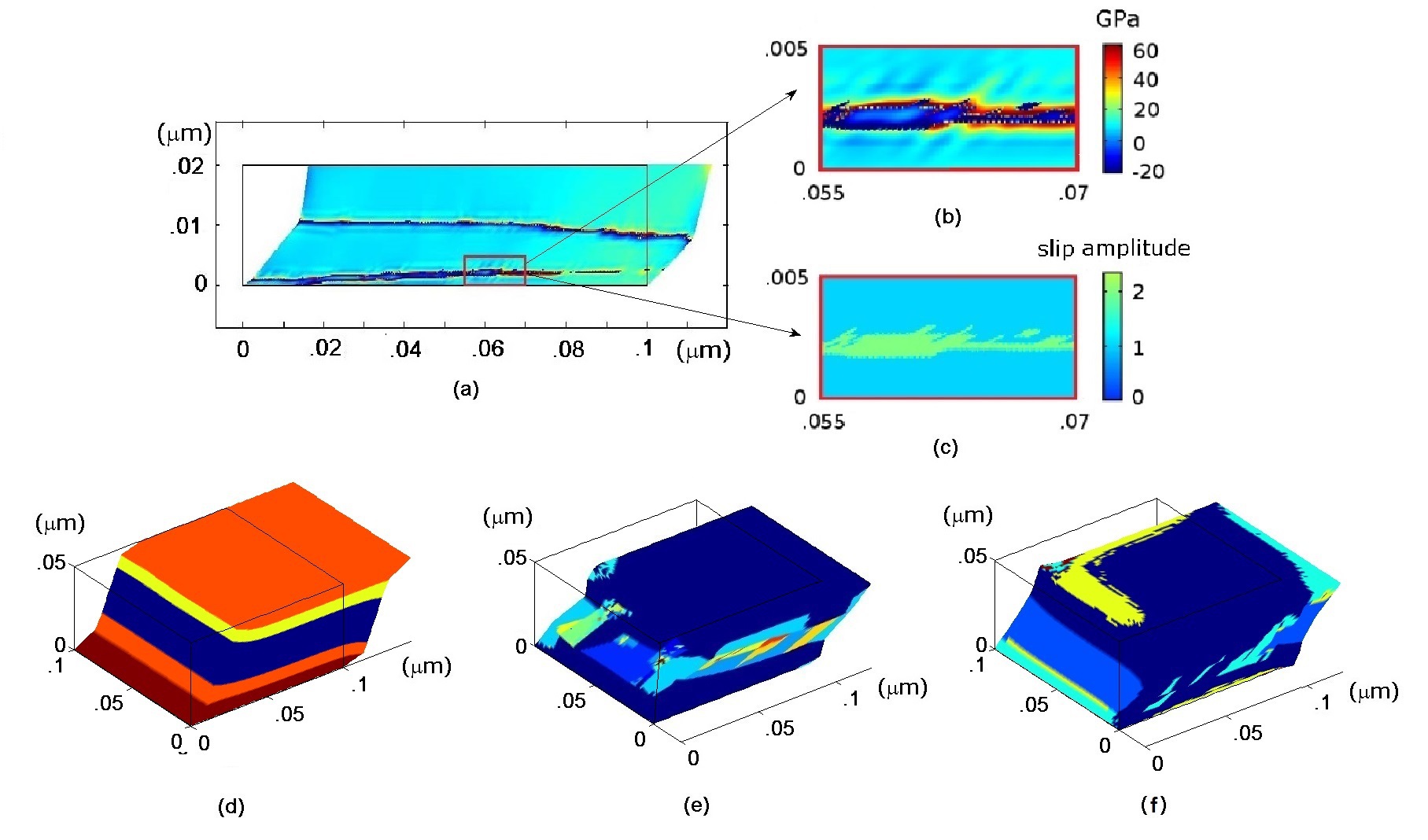}
\caption{(Color online) Microstructures developed in plastic flow. The thin black lines outline the undeformed reference configurations. (a) Band-like stress distribution during simulation I (section on the ($\mb{a},\mb{m}$)-plane). (b) Detail of the stress peaks at the highest plastification zones of a shear band, with (c) the corresponding slip amplitudes, where double slip has also occurred. (d) Almost-coherent shear bands involving compatible shears in simulation II. Dark blue is unsheared; the other colors indicate the principal slips activated by the loading, and different combinations thereof. (e) Complex microstructure during simulation III, which involves mostly kinematically compatible slips. (f) Complex microstructure developed in simulation V, which activates mostly non-compatible slips. A similar result to V holds for simulation IV (not shown), which also involves mostly non-compatible slips. Dark blue is unsheared; the other colored domains involve various combinations of the fundamental slips, forming kinematically compatible as well as non-compatible pairs.
}
\label{fig:FEPN}
\end{figure}

The results in Fig.~\ref{fig:FEPN_hyst}, showing the serrated stress-strain curves obtained from the numerical simulations, evidence the bursty character of slip-mediated plastic deformation in the model. The mechanical response of the crystal occurs through stress-relieving dislocation avalanches at many scales. Hardening is absent when only one or few compatible slips are activated as in simulations I, II, III, where, after the initial elastic climb the stress reaches a yielding value, after which it remains substantially at a plateau value, which depends on the number and compatibility properties of the involved slips. In contrast, hardening is present in the more general case of simulations IV and V (see the green and purple dashed lines in Fig.~\ref{fig:FEPN_hyst}, right), a typical effect in many experimental observations. As the initial crystal is fairly homogeneous, we observe in all cases the expected nucleation peak, common in other similar rate independent hysteretic phenomena, both theoretically and experimentally \cite{cot49,fed92,tru04,din12}. The static defects have a marked influence on the first-cycle peak, which significantly decreases with increasing quenched disorder (left inset in Fig.~\ref{fig:FEPN_hyst}). The peak is anyway lowered and eventually eliminated in subsequent loading cycles due to the dislocation structures developed during the previous cycles (see the blue line in the left panel, and the red dotted line in the right panel of Fig.~\ref{fig:FEPN_hyst}).

Fig.~\ref{fig:FEPN} demonstrates the crucial role that compatibility among the involved principal slips plays in the developing microstructures, and in controlling the creation of dislocations. Several features emerge, showing how the discrete plastic map $\FEPN$ captures many details of defect-mediated deformation. Panels (a)-(d), corresponding to loadings I and II, highlight the similarities in plastic processes which involve a single or several compatible slips. In both cases, at an intermediate yielding stage the plastic deformation is organized in bands which are in rough alignment with the optimal directions for stress relaxation at the $\FEPN$ discontinuities, with dislocations mostly concentrated at these non-fully-coherent interfaces. However, depending on the boundary conditions or the lattice orientation, much more complex microstructures may arise also in the case of activated slips which are mostly compatible, as happens in simulation III, see Fig.~\ref{fig:FEPN}(e).  Despite the complexity of the microstructure developed in simulation III, the fact that it involves mostly compatible slips is likely at the origin of the lack of hardening observed in this case in the material response. Complex microstructures are a fortiori created in simulations IV and V, wherein the more generic loading conditions activate a (sub)set of non-compatible slips, see Fig.~\ref{fig:FEPN}(e) showing simulation V as an example. In these more general loadings the non-compatible fronts involve strong defect nucleation, although the deforming lattice typically attempts to accommodate the boundary conditions through selective (cross-)slip combinations, generating compatible interfaces as much as permitted by local minimization. However, the evolving structures are unable to fully obviate to the energetic cost imposed by non-compatibility, and the ensuing rapid accumulation of internal stresses and defects is a likely factor \cite{04boce} in the hardening observed in the stress plateau of simulations IV and V, as in the green and purple dashed lines in Fig.~\ref{fig:FEPN_hyst} (right).

\begin{figure}
\centering
\includegraphics[width=.95\textwidth, clip, keepaspectratio]{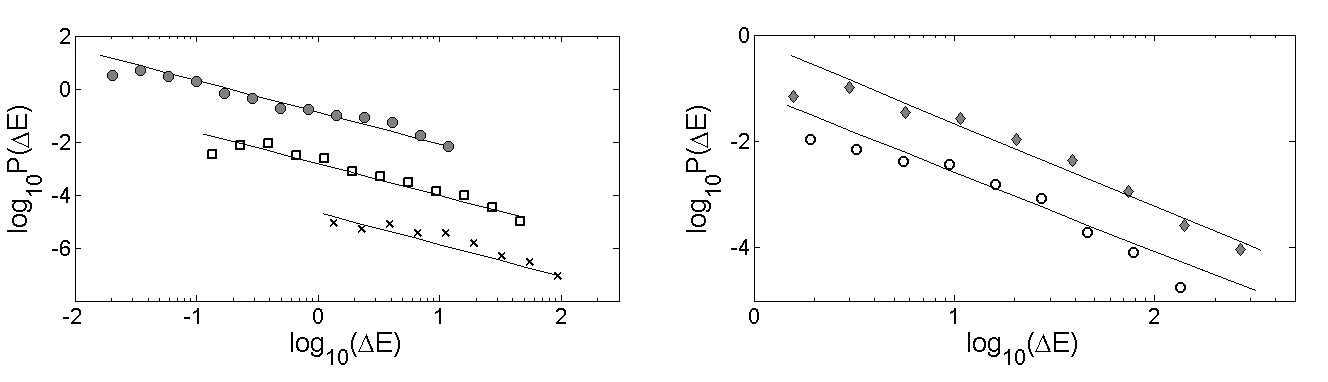}
\caption{Log-log plot of the heavy-tailed probability density of the energy drops along the plastic hysteresis (logarithmic binning). Left: data from simulations I (circles), II (squares), III (crosses). Right: data from simulations IV (diamonds) and V (circles).  Data for simulations II, III, and V, are shifted down for clarity.
The solid lines, drawn to guide the eye, indicate power laws with exponent $1.2$ (left) and $1.5$ (right). See also Table~\ref{tab1}.}
\label{fig:bin14b}
\end{figure}

{\bf Intermittency.~}Fig.~\ref{fig:bin14b} shows the statistical properties, derived from the energy drops accompanying the slip avalanches during plastification, which generate the spatial patterns in Fig.~\ref{fig:FEPN}. The three loading conditions I, II, III, in which only compatible shears are activated, produce largely equivalent heavy tailed statistics. We see that the size distributions associated to these fast dissipation events decay compatibly with power laws over about two decades, with exponents near or just below 1.2, see Table~\ref{tab1}. Simulation IV and V, on the other hand, produce heavy-tailed statistics consistent with a higher power-law exponent near $1.5$. Analogous results hold for the stress drops, with respective exponents near 1.5 and 1.7, as is also reported in Table~\ref{tab1}. Interestingly, the higher exponents pertain to the cases showing evidence, such as hardening, of a higher defect density induced by the imposed loads due to the kinematic non-compatibility of the activated slips. This might lead to dislocation entanglements possibly screening long-range defect interactions and acting as barriers to free dislocation glide, with, in turn, a reduction of avalanche size and an increase of the statistics' exponents, see also \cite{lev15}. Detecting such compatibility-related effects, which are known to play important roles also in the (reversible) martensitic transformations of crystalline substances such as shape-memory alloys \cite{13james_nat,bbbgz15}, is a main advantage of our method based on an exact, 3D multiplicative evolution law (\ref{updateFG}) for the plastic strain.

The above results on the dependence of the plasticity exponents on the specific wells involved in the flow, and thus, on the specific symmetry of the crystal and the specific loading imposed to the material, may be juxtaposed the analogous effects occurring in the martensitic transformations of crystalline materials such as shape-memory alloys. In the latter, it is known that scale-free behavior is characterized by exponents that are material- and loading-dependent, due to the symmetry, the compatibility properties, and the number, of the energy wells involved by the specific boundary conditions driving the transformation \cite{niemann12,niemann14,vives_planes_13,toth_14}. Our present results evidence analogous effects also in scale-free crystalline plastic flow.

\begin{table}[t]
    \begin{center}
    \begin{tabular}{| l | l | l | l |}\hline
    Simulation & Exponent energy drops & Exponent stress drops\\\hline
    \;\;I & \;\;\;\;\;\;$1.22 \pm 0.01$ & \;\;\;\;\;\;$1.46 \pm 0.03$\\\hline
    \;\;II & \;\;\;\;\;\;$1.19 \pm 0.01$ & \;\;\;\;\;\;$1.49 \pm 0.04$ \\\hline
    \;\;III & \;\;\;\;\;\;$1.15 \pm 0.02$ & \;\;\;\;\;\;$1.48 \pm 0.03$\\\hline
    \;\;IV & \;\;\;\;\;\;$1.54 \pm 0.05$ & \;\;\;\;\;\;$1.75 \pm 0.05$\\\hline
    \;\;V & \;\;\;\;\;\;$1.57 \pm 0.05$ & \;\;\;\;\;\;$1.72 \pm 0.04$\\\hline
    \end{tabular}
   \caption{Exponents, computed by the maximum likelihood method \cite{newman}, for the distributions of the energy and stress drops pertaining to simulations I-V. See also Fig.~\ref{fig:bin14b}.}
   \label{tab1}
    \end{center}
\end{table}

\section{Conclusions}\label{sec:disc}

We have developed a nonlinear 3D phase field model accounting for the infinite and discrete invariance group of periodic lattices, which captures the main significant effects in crystal plasticity at microscopic scales. The non-compatible plastic distortion jumps resolve the single lattice defects, and plasticity in the model spontaneously proceeds through intermittent defect nucleation and evolution in the lattice, with jump fronts orienting themselves to reduce the elastic energy, and with the plastic distortion always driven by the phase fields towards crystallographically-preferred configurations. Via the relation $G =  E^{-1}\GLZ  E$ earlier recalled, and by adopting the appropriate set of principal slips $\Spm_\alpha$, the present approach can be adapted to investigate such phenomena also in crystalline materials whose symmetry is other than simple cubic as considered above.

The numerical results show how both simple and complex strain topologies may develop in the deformed crystal, depending on the external loading and the kinematic compatibility properties of the different slip systems that are activated. Simulated plastic flow also shows a power-law decay for the avalanching events. The simplest cases mostly involving compatible active slips exhibit band-structured plastic deformation, low dislocation density and no hardening. For these we estimate low exponents agreeing with the low exponents recently determined in the detailed DDD simulations of plastic yielding \cite{zapperi13} and in the re-analysis in \cite{zha12,gu13} of  experimental microplasticity data in \cite{dim06,08bri,uch09}. We find higher exponents in more generic plastification conditions, when the subset of activated slips are mostly non-compatible, producing complex microstructures and intermittent yielding accompanied by hardening. These results suggest that microscale crystal plasticity should not belong \cite{zapperi13} to the universality class characterized by exponents compatible with mean-field depinning \cite{zai06,csi07,tse13}. In general, scale-free plastic flow in crystals may actually be material and loading dependent and lack universality, as is also advanced in recent work \cite{lev15,zapperi_review_14}, and in possible analogy to the material and loading dependent statistical features of avalanching behavior also observed in reversible martensitic transformation in crystals \cite{niemann12,niemann14,vives_planes_13,toth_14}.

\bigskip

\noindent{\it Acknowledgements.~}We acknowledge financial support from the Italian PRIN Contract 200959L72B004 and from a SAES Getters - Politecnico di Milano research contract.

\end{document}